\documentclass[aps,pra,superscriptaddress]{revtex4}

\usepackage[dvips]{graphicx}
\usepackage{indentfirst}

\def\be{\begin{equation}}
\def\ee{\end{equation}}
\def\bea{\begin{eqnarray}}
\def\eea{\end{eqnarray}}
\newcommand{\ket}[1]{\mbox{$|#1\rangle$}}
\newcommand{\bra}[1]{\mbox{$\langle#1|$}}
\def\za{\sigma_a^z}
\def\zb{\sigma_b^z}
\def\zj{\sigma_j^z}
\def\pa{\sigma_a^+}
\def\pb{\sigma_b^+}
\def\pj{\sigma_j^+}
\def\ma{\sigma_a^-}
\def\mb{\sigma_b^-}
\def\mj{\sigma_j^-}
\def\ke{{\mathbf{k},\mathbf{\epsilon}}}
\def\kep{{\mathbf{k}^{\prime},\mathbf{\epsilon}^{\prime}}}

\newcommand\ddtavg[1]{\frac{\partial{\langle}#1{\rangle}}{{\partial}t}}
\newcommand\avg[1]{{\langle}#1{\rangle}}

\begin{document}
\title{Controlling dipole-dipole frequency shifts in a lattice-based optical atomic clock}

\author{D.E. Chang}
\email{dechang@fas.harvard.edu} \address{Department of Physics,
Harvard University, Cambridge, Massachusetts 02138}
\author{Jun Ye}
\address{JILA, National Institute of Standards and Technology and
University of Colorado, Boulder, Colorado 80309-0440}
\author{M.D. Lukin}
\address{Department of Physics, Harvard University, Cambridge,
Massachusetts 02138}

\date{\today}

\begin{abstract}
Motivated by the ideas of using cold alkaline earth atoms trapped
in an optical lattice for realization of optical atomic clocks, we
investigate theoretically the perturbative effects of atom-atom
interactions on a clock transition frequency.  These interactions
are mediated by the dipole fields associated with the optically
excited atoms.  We predict resonance-like features in the
frequency shifts when constructive interference among atomic
dipoles occur. We theoretically demonstrate that by fine-tuning
the coherent dipole-dipole couplings in appropriately designed
lattice geometries, the undesirable frequency shifts can be
greatly suppressed.
\end{abstract}

\maketitle
%
\section{Introduction}
The development of increasingly accurate atomic clocks has led to
many advances in technology and tests of fundamental physics. In
the search for the next generation of clocks and frequency
standards, there has been considerable interest in using alkaline
earth species because of their narrow intercombination lines in
the optical spectrum~\cite{hall89}.  In order to achieve a high
level of short-term stability and long-term reproducibility and
accuracy on the clock transition, it is desirable to have a large
number of cold atoms located in a well-characterized trap for an
improved signal-to-noise ratio ($S/N$) and reduced systematic
errors associated with atomic motion.  Single ion-based systems do
effectively eliminate Doppler and other motion-related systematic
errors when the single ions are confined in the Lamb-Dicke
regime~\cite{rafac00}, although the achievable $S/N$ is limited by
single quantum absorbers.  For neutral atoms it is important that
changes in the level structure due to the trapping potential do
not alter the relevant clock transition frequency. Such a scheme
has been proposed by trapping alkaline earth atoms in
three-dimensional optical lattices tuned to a ``magic" wavelength
where the relevant states for the clock transition experience
exactly the same level shift~\cite{katori02}. The
${}^{1}S_0(F=9/2)-{}^{3}P_0(F=9/2)$ forbidden transition
($\lambda_0=700$ nm) in ${}^{87}$Sr~\cite{katori02} is in
particular a promising candidate for a lattice-based optical clock
transition because of the long lifetime of the excited state
(${\sim}160$ s) and the insensitivity of the $J=0$ states to the
polarization state of the trapping light. Already there have been
efforts towards the cooling and trapping of
${}^{87}$Sr~\cite{mukaiyama03,xu03,xu03josab}, and recently this
transition was directly observed and measured for the first
time~\cite{courtillot03}.  Calcium, another alkaline earth atom
that has been studied extensively as a frequency
standard~\cite{Udem01,Wilpers02}, may be a candidate for optical
lattice clocks as well.

In the case of $N$ independent atoms, one benefits from a
$\sqrt{N}$ improvement in $S/N$ in spectroscopy. However, atoms
trapped in an optical lattice can interact with each other and
cannot truly be considered independent.  Each optically excited
atom represents essentially a point dipole whose radiated
electromagnetic field can affect other atoms.  These atom-atom
interactions can manifest themselves as shifts in the observed
transition frequencies. Because of the spatial ordering of atoms
in a lattice and the potentially high atomic density, it is
possible that such interactions may produce very large frequency
shifts. One might expect then that dipole-dipole interactions can
be much more severe here than in, for example, atomic fountains,
and thus could place serious limits on the accuracy of an optical
lattice clock if not accounted for. On the other hand, it might be
possible to design lattice geometries where this shift is reduced
or cancelled. Although the trapping lasers are constrained to
operate at the ``magic" wavelength, the lattice geometry can be
altered by changing the relative orientations of the trapping
beams, whose degrees of freedom can be characterized by a set of
variables $\{\alpha\}$.

In this paper, we investigate theoretically the dipole-dipole
interaction-induced shifts in the clock transition frequency
recovered by Ramsey spectroscopy. We show that by varying the
lattice geometry we can quantitatively control the clock frequency
shift and even reduce the shift to zero.  In particular, we give
an analytical equation that can be solved giving configurations
$\{\alpha_0\}$ where constructive interference causes the line
shift to be very large. In these ``bad" lattice configurations,
the magnitude of the line shift scales approximately like
$N^{2/3}$. Quite generally we propose that by tuning the parameter
space $\{\alpha\}$ to lie in between two of these bad
configurations, one can find ``good" configurations where the
shift is cancelled. The mechanism of cancellation is associated
with the destructive interference of contributions to the shift
from different atoms in the lattice.

It is important to emphasize that the present mechanism and
theoretical treatment differ considerably from the conventional
approaches used to treat dipole-dipole line broadening and shifts.
In the case of atoms in a spatially-ordered lattice geometry,
long-range effects are important, and the usual methods involving
binary collisions of nearest neighbors~\cite{allard82} are not
applicable.  These long-range effects include, in particular,
interference of the far-field dipole radiation produced by the
excited atoms, a phenomenon similar to Bragg scattering in a
crystal.

This paper is organized as follows. In Sec.~\ref{sec:eqofmotion}
we derive equations describing the evolution of an atomic system
with dipole-dipole interactions. These equations are derived
assuming that the atoms are in the Lamb-Dicke regime, with one
atom or less per lattice site.  In Sec.~\ref{sec:ramsey} we give a
brief review of Ramsey spectroscopy and solve for the
dipole-dipole induced line shift using perturbation theory.  We
find that the shift can be qualitatively understood in terms of
the classical interaction energies between oscillating dipoles.
There is a contribution to the shift that is zeroth order in the
interrogation time $t$, which is due to imperfections in the
Ramsey pulses.  Even with perfect pulses, however, one finds a
shift that is first order in $t$ that results from spontaneous
decay of the atoms. Sec.~\ref{sec:generalization} discusses how
our result for the line shift can be generalized for systems with
imperfect filling of the lattice sites and for multilevel atoms.
In the case of imperfect filling, one can calculate the mean value
of the frequency shift as well as some nonzero variance, due to
the uncertainty of how the lattice is filled. In
Sec.~\ref{sec:latticedesign}, we derive an equation that can be
solved giving lattice configurations where the shift is large due
to constructive interference.  We derive an approximate scaling
law for the shift in these ``bad" configurations and discuss how
the line shift can be reduced by choosing an appropriate lattice
design. In Sec.~\ref{sec:example} we demonstrate these results
numerically for one specific lattice configuration.

\section{Equations of motion}\label{sec:eqofmotion}
To treat the problem of interacting atoms in a lattice, we
consider $N$ two-level atoms in the Lamb-Dicke limit with
polarizability along the $z$-axis.  A simple model of the system
consists of treating the atoms as point dipoles, and we further
assume that there is one or less atom per lattice site. This
corresponds to a Mott-insulator state for bosons or a normal state
for fermions.  In principle, to solve exactly the problem of
interacting atoms one would start from the full atom-field
Hamiltonian and take into account not only all the atomic degrees
of freedom but the continuum of electromagnetic field modes.  To
simplify the theoretical treatment, we effectively eliminate the
field in the standard way using the Born-Markov approximation~(see
Appendix). This is valid provided that the atomic system evolves
slowly on timescales of the correlation time $\tau_c$, which is of
the order $L/c$ where $L$ is the linear size of the system and $c$
is the speed of light. As a result of eliminating the field, one
finds an effective equation of motion for the density matrix
$\rho$ of the atomic system. Atom-atom interactions then appear
through an effective Hamiltonian $H_{\scriptsize{\textrm{eff}}}$
as well as through a non-Hermitian operator $\mathcal{L}$:
\be \frac{\partial\rho}{{\partial}t}=
\frac{1}{i\hbar}\left[H_0+H_{\scriptsize{\textrm{eff}}},\rho\right]+\mathcal{L}[\rho].
\ee
Here, $H_0$ is the atomic Hamiltonian for a non-interacting
system.  Writing out all the terms in detail,
\bea \frac{\partial\rho}{{\partial}t} & = & \frac{1}{i}\sum^{}_{a}
\left[\frac{\omega_0}{2}\za,\rho\right]
-\frac{i\Gamma}{2}\sum^{}_{a{\neq}b}g(k\mathbf{r}_{ab})\left[\pa\mb,\rho\right] \nonumber\\
& &
\label{eq:dpdt}-\frac{\Gamma}{2}\sum^{}_{a,b}f(k\mathbf{r}_{ab})\left(\left\{\pa\mb,\rho\right\}-2\mb\rho\pa\right)
-\frac{1}{4}\sum^{}_{a}\gamma\left(\rho-\za\rho\za\right), \eea
where
\bea f(\mathbf{v}) & = &
\frac{3}{2}\left(\sin^{2}\theta\frac{\sin{v}}{v}+(3\cos^{2}\theta-1)\left(\frac{\sin{v}}{v^3}-\frac{\cos{v}}{v^2}\right)\right),
\nonumber \\
\label{eq:fg}g(\mathbf{v})& = &
-\frac{3}{2}\left(\sin^{2}\theta\frac{\cos{v}}{v}+(3\cos^{2}\theta-1)\left(\frac{\cos{v}}{v^3}+\frac{\sin{v}}{v^2}\right)\right),
\eea
and $\theta$ is the angle that $\mathbf{v}$ makes with the
$z$-axis.

The first term on the right-hand side of Eq.~(\ref{eq:dpdt})
corresponds to $H_0$.  $\za$ is the Pauli matrix of atom $a$
corresponding to the population difference between the excited and
ground states, and $\omega_0$ is the resonance frequency of the
dipole transition. The second term corresponds to
$H_{\scriptsize{\textrm{eff}}}$.
 Here, $\Gamma=k_{0}^{3}d^2/3\pi\epsilon_{0}\hbar$ is the spontaneous
decay rate of the excited state of a single, isolated atom, where
$k_{0}=2\pi/\lambda_0=\omega_0/c$ and $d$ is the dipole matrix
element between the ground and excited states. $\pa$ is the atomic
raising operator on atom $a$, and $\mb$ is the lowering operator
on atom $b$.  One then sees that the effect of dipole-dipole
interactions is an exchange of excitation between pairs of atoms.
The strength of interaction is modified by a function
$g(k\mathbf{r}_{ab})$ that depends on the distance and orientation
between two dipoles. It is to be understood that $k=k_0$ in the
functions $f$ and $g$. We see that both short-range,
near-field~$(1/r^3)$ and long-range, far-field~$(1/r)$ dipole
interactions are included in our formalism and are treated on
equal footing. The third term on the right side of
Eq.~(\ref{eq:dpdt}) corresponds to $\mathcal{L}$ and also is due
to atom-atom interactions. It also depends on $\Gamma$ and has a
position dependence described by $f(k\mathbf{r}_{ab})$. The
non-Hermitian nature of this term is evident through the
anticommutator. Physically $\mathcal{L}$ describes the processes
of both independent and cooperative decay. Finally, we have also
added phenomenological dephasing through the $\gamma$ term, which
in particular includes the effects of a finite, short-term
linewidth of the laser interrogating the clock transition.

From Eq.~(\ref{eq:dpdt}), one can derive equations of motion for
any atomic operators. For our particular application of Ramsey
spectroscopy, we find it necessary to solve for the coherence
$\avg{\pa}$ and the two-atom correlation $\avg{\za\pb}$.
Furthermore, to remove the rapid oscillations due to $\omega_0$,
it is convenient to work in the frame rotating with the
interrogating laser frequency $\omega_L$, where
\bea \ddtavg{\pa} & = &
-i\delta\avg{\pa}-\frac{\Gamma+\gamma}{2}\avg{\pa}
+\frac{\Gamma}{2}\sum^{}_{b{\neq}a}\left(f(k\mathbf{r}_{ab})-ig(k\mathbf{r}_{ab})\right)\avg{\za\pb}\label{eq:ddtpa},
\\
\ddtavg{\za\pb} & = &
-\left(i\delta+\frac{3\Gamma+\gamma}{2}\right)\avg{\za\pb}-\Gamma\avg{\pb}
-\frac{\Gamma}{2}\left(f(k\mathbf{r}_{ab})+ig(k\mathbf{r}_{ab})\right)\avg{\pa}
-{\Gamma}f(k\mathbf{r}_{ab})\avg{\pa\zb} \nonumber \\ & &
+\frac{\Gamma}{2}\sum^{}_{j{\neq}a,b}\left(f(k\mathbf{r}_{bj})-ig(k\mathbf{r}_{bj})\right)\avg{\za\zb\pj}
-\Gamma\sum^{}_{j{\neq}a,b}\left(f(k\mathbf{r}_{aj})+ig(k\mathbf{r}_{aj})\right)\avg{\pa\pb\mj}
\nonumber \\ & &
-\Gamma\sum^{}_{j{\neq}a,b}\left(f(k\mathbf{r}_{aj})-ig(k\mathbf{r}_{aj})\right)\avg{\ma\pb\pj}\label{eq:ddtzapb},
\eea
and $\delta=\omega_L-\omega_0$.

In principle, to solve for the atomic system exactly, equations of
motion for higher-order correlations are needed, although
typically some approximation is used to truncate the resulting
hierarchy of equations.  In general these equations can describe
light-matter interactions in an optically dense medium, including
radiation trapping, level shifts, and
superradiance~\cite{fleischhauer99}.

\section{Ramsey spectroscopy in optical lattice clock}\label{sec:ramsey}
\subsection{Basic principles}
We now analyze the effects of atom-atom interactions on Ramsey
spectroscopy.  Starting from the ground state
$\ket{g}^{{\otimes}N}$ of the system, suppose that one applies a
strong probe pulse with the interrogating laser, given in the
rotating frame by the Hamiltonian
\be
H=\sum^{}_{a}i\hbar\Omega\left({\pa}e^{i\mathbf{k}{\cdot}\mathbf{r}_a}-{\ma}e^{-i\mathbf{k}{\cdot}\mathbf{r}_a}\right),
\ee
where $\Omega$ is the Rabi frequency. For simplicity, we have made
a plane-wave assumption about the probing laser, taking
$\mathbf{k}$ to be in the positive $x$ direction. We also assume
that $k{\approx}k_0$, and suppress the subscript in future
calculations.  We can do this if the phase error
$e^{i{\delta}L/c}$ over the length of the sample incurred by
making this assumption is small. Applying this pulse for a time
$\tau$ evolves the system through the unitary operator
\be U={\prod^{}_{a}}^{\otimes}\left(\begin{array}{cc} \cos\Omega\tau & e^{ikx_a}\sin\Omega\tau \\
-e^{-ikx_a}\sin\Omega\tau & \cos\Omega\tau
\end{array}\right).
\ee
The state vector immediately following this pulse is given by
\be
\ket{\psi_{i}}={\prod^{}_{a}}^{\otimes}\left(\cos\Omega\tau\ket{g}+e^{ikx_a}\sin\Omega\tau\ket{e}\right),
\ee
where $\ket{e}$ denotes the excited state of the clock transition.

In Ramsey spectroscopy, one lets $\ket{\psi_i}$ evolve for an
interrogation time $t$ to $\rho(t)$.  During this time the
coherence between the ground and excited states acquires some
time-dependent phase that depends on $\delta$. After a time $t$,
one then applies a second pulse corresponding to the inverse
unitary operation $U^{\dagger}$, and then measures the signal
corresponding to $\tilde{S}=\sum^{}_{a}\za$, averaged over the
final system $\rho_f$. $\tilde{S}$ corresponds to the total
population inversion. Because of the second pulse, $\tilde{S}$
will now depend on $\delta$ and $t$, allowing one to extract
information about the resonance line.

Formally, we can rewrite $\tilde{S}$ as
\bea \tilde{S} & = & \textrm{Tr}\left(\sum^{}_{a}{\za}U^{\dagger}\rho(t)U\right) \nonumber \\
& = & \cos{2\Omega\tau}\avg{\sum^{}_{a}\za}
-2\sin{2\Omega\tau}\textrm{Re}\avg{\sum^{}_{a}e^{ikx_a}\pa}\label{eq:signaltilde},
\eea
where the averages denoted above apply to $\rho(t)$, the system
immediately before the second pulse.

In the case of $N$ non-interacting, independently decaying atoms,
\bea \avg{\za} & = &
-1+e^{-{\Gamma}t}\left(1-\cos{2\Omega\tau}\right), \\
\avg{e^{ikx_a}\pa} & = &
\frac{1}{2}\sin{2\Omega\tau}e^{-(i\delta+\Gamma/2+\gamma/2)t},
\eea
which gives a corresponding signal
\be
\label{eq:independentsignal}\tilde{S}=-N\left(\cos{2\Omega\tau}\left(1-e^{-{\Gamma}t}\right)+e^{-{\Gamma}t}\cos^{2}{2\Omega\tau}+e^{-({\Gamma}+\gamma)t/2}\sin^{2}{2\Omega\tau}\cos{{\delta}t}\right).
\ee
One can see that there is a peak in the signal around $\delta=0$.
Determination of this peak allows one to find the frequency of the
transition.  One can note two important points about $\tilde{S}$.
The contrast in $\tilde{S}$ with respect to $\delta$ is maximized
when a ``perfect" $\pi/2$ pulse is applied, \textit{i.e.}, when
$\Omega\tau=\pi/4$.  Furthermore, the contribution to $\tilde{S}$
due to $\avg{\za}$ in Eq.~(\ref{eq:signaltilde}) is independent of
$\delta$ and thus plays no role in determination of the resonance
line.  Thus, one is motivated to define an effective signal $S$
that consists of the part of $\tilde{S}$ that is actually used to
determine the line:
\be\label{eq:signal}
S=-2\sin{2\Omega\tau}\textrm{Re}\avg{\sum^{}_{a}e^{ikx_a}\pa}. \ee
The equation above states that from a theoretical standpoint,
determination of the resonance line by measuring the population
inversion after the second Ramsey pulse is equivalent to measuring
the real part of $\avg{e^{ikx_a}\pa}$ directly before the second
pulse.
%
\subsection{Effect of interactions}
Solving for $S$ exactly in the presence of dipole-dipole
interactions appears to be quite a difficult task.  Since all
interactions are proportional to $\Gamma$, our approach is to
solve for $S$ as a perturbative expansion in ${\Gamma}$.  In
particular, we solve for the coherence $\avg{\pa}$ in
Eq.~(\ref{eq:ddtpa}) to second order in $\Gamma$.  This requires
solving $\avg{\za\pb}$ in Eq.~(\ref{eq:ddtzapb}) to first order in
$\Gamma$. The solution for the coherence with appropriate initial
conditions is
\bea \avg{\pa} & = &
\frac{1}{2}e^{-ikx_a}\sin(2\Omega\tau)e^{-\left(i\delta+\gamma/2\right)t}\left[1-\frac{{\Gamma}t}{2}
(1+C_a)+\frac{({\Gamma}t)^2}{8}(1+C_a)\right. \nonumber \\
& &
+\left.\frac{\Gamma^2}{2\gamma^2}\sum^{}_{b{\neq}a}\left(f(k\mathbf{r}_{ab})-ig(k\mathbf{r}_{ab})\right)
\left(A_{ab}\gamma^{2}t^{2}-2B_{ab}\left(e^{-{\gamma}t}+{\gamma}t-1\right)\right)\right]\label{eq:pafull},
\eea
where
\bea A_{ab} & = &
-\frac{3}{4}e^{ik(x_a-x_b)}\cos(2\Omega\tau)+\frac{1}{2}e^{ik(x_a-x_b)}+\frac{1}{4}\left(f(k\mathbf{r}_{ab})+ig(k\mathbf{r}_{ab})\right)
-\frac{1}{2}f(k\mathbf{r}_{ab})\cos(2\Omega\tau) \nonumber \\
& &
-\frac{1}{4}\sum^{}_{j{\neq}a,b}\left(f(k\mathbf{r}_{bj})-ig(k\mathbf{r}_{bj})\right)e^{ik(x_a-x_j)}\cos^{2}(2\Omega\tau),
\\
B_{ab} & = &
\frac{1}{4}e^{ik(x_a-x_b)}\sin^{2}(2\Omega\tau)\sum^{}_{j{\neq}a,b}\left(f(k\mathbf{r}_{aj})\cos{kx_{aj}}+g(k\mathbf{r}_{aj})\sin{kx_{aj}}\right),
\\
C_{a} & = &
\sum^{}_{b{\neq}a}\left(f(k\mathbf{r}_{ab})-ig(k\mathbf{r}_{ab})\right)e^{ik(x_a-x_b)}\cos{2\Omega\tau}.
\eea
Eq.~(\ref{eq:pafull}) is correct to every order of $\gamma$.
Eqs.~(\ref{eq:signal}) and (\ref{eq:pafull}) can be evaluated
numerically for a given lattice configuration and number of atoms.
To illustrate the general features of the shift, however, we now
make the following simplifications.  We expand
Eq.~(\ref{eq:pafull}) to lowest order in $\gamma$.  We also assume
that the Ramsey pulses are nearly perfect $\pi/2$ pulses,
\textit{i.e.}, $\cos{2\Omega\tau}=\epsilon{\ll}1$. We then keep
terms like $\epsilon{\Gamma}t$ but ignore terms like
$\epsilon\Gamma^{2}t^2.$ With these simplifications,
\be\label{eq:pa}
\avg{\pa}{\approx}\frac{1}{2}e^{-ikx_a}\sin(2\Omega\tau)e^{-i{\delta}t}\left(1-\frac{{\Gamma}t}{2}
+\frac{({\Gamma}t)^2}{8}-\phi_{a}\right), \ee
where
\bea \phi_{a} & = &
\sum^{}_{b{\neq}a}\left(f(k\mathbf{r}_{ab})-ig(k\mathbf{r}_{ab})\right)e^{ik(x_a-x_b)}
\left(\frac{{\Gamma}t}{2}\cos(2\Omega\tau)+\frac{({\Gamma}t)^2}{4}
\left(1+\frac{1}{2}\left(f(k\mathbf{r}_{ab})+ig(k\mathbf{r}_{ab})\right)e^{-ik(x_a-x_b)}\right.\right.
\nonumber \\ & &
+\left.\left.\frac{1}{2}\sum^{}_{j{\neq}a,b}\left(f(k\mathbf{r}_{aj})\cos{kx_{aj}}+g(k\mathbf{r}_{aj})\sin{kx_{aj}}\right)\right)\right)
\eea
The first three terms in the parentheses of Eq.~(\ref{eq:pa}) are
a result of expanding the $e^{-{\Gamma}t/2}$ term that appears in
the result for independent atoms, given in
Eq.~(\ref{eq:independentsignal}). This is just the decay of the
signal one would get from independent spontaneous emission. The
last term in the parentheses is a correction due to atom-atom
interactions. Plugging this result into Eq.~(\ref{eq:signal}), we
find that
\be
S{\approx}-\sin^{2}{2\Omega\tau}\left((\cos{\delta}t)\left(N-\frac{N{\Gamma}t}{2}
+\frac{N({\Gamma}t)^2}{8}-\sum^{}_a\textrm{Re}\phi_a\right)
-(\sin{\delta}t)\sum^{}_{a}\textrm{Im}\phi_a\right). \ee
Because of the antisymmetric $\sin{\delta}t$ term now appearing in
$S$, one immediately sees that dipole-dipole interactions
introduce a shift $\delta_p$ in the Ramsey fringes, which can be
found by solving ${\partial}S/{\partial}\delta=0$. Suppose that
the inequalities $\delta_{p}t{\ll}1,{\Gamma}t{\ll}1$ are
satisfied. Under these conditions, a simple expression for
$\delta_p$ results:
\bea \frac{\delta_p}{\Gamma} & \approx &
\frac{1}{N}\sum^{}_{a}\sum^{}_{b{\neq}a}\left(g(k\mathbf{r}_{ab})\cos{kx_{ab}}-f(k\mathbf{r}_{ab})\sin{kx_{ab}}\right)\times
\nonumber \\ & &
\left(\frac{1}{2}\cos{2\Omega\tau}+\frac{{\Gamma}t}{4}
\left(1+\frac{1}{2}\sum^{}_{j{\neq}a,b}\left(f(k\mathbf{r}_{aj})\cos{kx_{aj}}+g(k\mathbf{r}_{aj})\sin{kx_{aj}}\right)\right)\right)\label{eq:shift}.
\eea
%
\subsection{Interpretation of shift}
The shift given by Eq.~(\ref{eq:shift}) yields a simple
interpretation.  In anticipation of future analysis, we write
$\delta_p$ as
\be\label{eq:shiftsymbolic}
\frac{\delta_p}{\Gamma}=\frac{1}{N}\sum^{}_{a}\sum^{}_{b{\neq}a}\tilde{U}_{ab}\times
\left(\frac{1}{2}\epsilon+\frac{{\Gamma}t}{4}\tilde{\Gamma}_{a}\right),
\ee
where
\bea \tilde{U}_{ab} & = &
g(k\mathbf{r}_{ab})\cos{kx_{ab}}-f(k\mathbf{r}_{ab})\sin{kx_{ab}},
\\ \epsilon & = & \cos{2\Omega\tau}, \\
\tilde{\Gamma}_{a} & = &
1+\frac{1}{2}\sum^{}_{j{\neq}a,b}\left(f(k\mathbf{r}_{aj})\cos{kx_{aj}}+g(k\mathbf{r}_{aj})\sin{kx_{aj}}\right).
\eea
We will see that $\tilde{U}_{ab}$ is a dimensionless quantity
corresponding to the classical interaction energy between two
oscillating dipoles, $\epsilon$ is a parameter characterizing the
error in the Ramsey pulses, and $\tilde{\Gamma}_{a}$ is a
dimensionless quantity characterizing cooperative decay of the
system.

To show the meaning of the $\tilde{U}_{ab}$ term in
Eq.~(\ref{eq:shiftsymbolic}), consider the interaction between a
classical, oscillating dipole at $\mathbf{r}_a$ excited with phase
$e^{i(kx_a-{\omega}t)}$ and the field incident on it due to a
classical, oscillating dipole at $\mathbf{r}_b$ excited with phase
$e^{i(kx_b-{\omega}t)}$.  We assume that both dipoles are oriented
along the $z$-axis and that their magnitudes $d$ are determined
from the relation $\Gamma=k_{0}^{3}d^{2}/3\pi\epsilon_{0}\hbar$.
The classical interaction energy between dipole $a$ and the
incident field is given by
$U_{ab}=-(1/2)\textrm{Re}(\mathbf{d}_{a}{\cdot}\mathbf{E}^{*}_b(\mathbf{r}_a))$.
The field at $\mathbf{r}_a$ due to dipole $b$ is~\cite{jackson99}:
\be\label{eq:classicalfield}
E_z(\mathbf{r}_a)=e^{i(kx_b-{\omega}t)}\frac{k^{3}d}{4\pi\epsilon_0}e^{ikr}\left(\frac{\sin^{2}\theta}{kr}
+(3\cos^{2}\theta-1)\left(\frac{1}{(kr)^3}-\frac{i}{(kr)^2}\right)\right),
\ee
where $r=|\mathbf{r}_b-\mathbf{r}_a|$.  Now using the definitions
in Eq.~(\ref{eq:fg}), the interaction energy can readily be
rewritten as:
\bea U_{ab} & = &
-\frac{1}{2}\textrm{Re}(\mathbf{d}_{a}\cdot\mathbf{E}_b^{*}(\mathbf{r}_a))=\frac{\hbar\Gamma}{4}\left(g(k\mathbf{r}_{ab})\cos{kx_{ab}}-f(k\mathbf{r}_{ab})\sin{kx_{ab}}\right)
\\ & = & \frac{\hbar\Gamma}{4}\tilde{U}_{ab}. \eea
One then sees that this indeed corresponds to the first term of
Eq.~(\ref{eq:shiftsymbolic}).

Although the $\tilde{U}_{ab}$ term in Eq.~(\ref{eq:shiftsymbolic})
resembles a classical interaction energy, the terms in parentheses
reflect the quantum mechanical nature of the system.  There is a
contribution to the shift that is zeroth order in the
interrogation time $t$ and proportional to $\epsilon$.  One notes
that for perfect $\pi/2$ Ramsey pulses, $\epsilon=0$. Thus, the
zeroth order shift is due to error in the Ramsey pulse.  This can
be understood by considering Eq.~(\ref{eq:ddtpa}). One sees that
the interaction terms only influence evolution of the coherence
through the term $\avg{\za\pb}$.  For a perfect $\pi/2$ pulse,
this term is initially zero, and in this case, interactions cannot
affect the measurement at short times.  This effect is due to the
nature of dipole-dipole interactions: these interactions cannot
influence the coherence $\avg{\pa}$ of an atom when it is in an
equal superposition of the ground and excited states.

Even if a perfect $\pi/2$ pulse is applied, there is an additional
contribution to the shift that is first order in $t$ and whose
strength is given by $\tilde{\Gamma}_{a}$. The intuition behind
this is also straightforward. Even if $\avg{\za\pb}$ is initially
zero, decay of the excited state will eventually evolve
$\avg{\za}$ away from zero and back towards its equilibrium value
of $-1$.  Once $\avg{\za\pb}$ is nonzero, interactions can
influence evolution of $\avg{\pa}$. The rate of decay is
characterized by $\tilde{\Gamma}_{a}$. The first term in
$\tilde{\Gamma}_{a}$ is the contribution from independent decay of
the atom back to the ground state. The second contribution
involves a sum over other atoms and represents a correction due to
the fact that the decay process may in fact be cooperative (e.g.,
superradiance). One can easily verify that the contribution from
atom $j$ is proportional to
$\textrm{Im}(\mathbf{d}_{a}\cdot\mathbf{E}_j^{*}(\mathbf{r}_a))$:
\be
\textrm{Im}(\mathbf{d}_{a}\cdot\mathbf{E}_j^{*}(\mathbf{r}_a)){\propto}f(k\mathbf{r}_{aj})\cos(kx_{aj})
+g(k\mathbf{r}_{aj})\sin(kx_{aj}). \ee
This reflects the well-known result that the atomic inversion
$\avg{\za}$ is driven by the dipole component in quadrature with
the incident field.
%
\section{Generalization of results}\label{sec:generalization}
\subsection{Imperfect filling of lattice sites}
Experimentally, knowing the exact number of atoms in the lattice
and achieving a filling factor of one atom per lattice site are
difficult tasks.  Most likely, one can experimentally determine
the density $\rho(\mathbf{r})$ of atoms in the lattice, such that
the probability of occupation at any particular site $a$ is
$P(\mathbf{r}_a)=\rho(\mathbf{r}_a)V$, where $V$ is the volume of
a unit cell.  It is straightforward to modify Eq.~(\ref{eq:shift})
to the case of imperfect filling. For simplicity, we only consider
the shift that is zeroth order in $t$.  This shift can be written
\bea \frac{\delta_p}{\Gamma} & = &
\frac{1}{N}\sum^{}_{a}\sum^{}_{b{\neq}a}\frac{1}{2}\cos{2\Omega\tau}
\left(g(k\mathbf{r}_{ab})\cos{kx_{ab}}-f(k\mathbf{r}_{ab})\sin{kx_{ab}}\right) \\
& = &
\frac{2}{N}\sum^{}_{\scriptsize{\textrm{pairs}}}\frac{1}{2}\cos{2\Omega\tau}\;g(k\mathbf{r}_{ab})\cos{kx_{ab}}
\\ & = &
\frac{1}{2}\cos{2\Omega\tau}\sum^{}_{\mathbf{R}{\neq}0}U(\mathbf{R})\frac{N(\mathbf{R})}{N},
\eea
where $U(\mathbf{R})=g(k\mathbf{R})\cos{kR_x}$, $\{\mathbf{R}\}$
denotes the set of direct lattice vectors, and $N(\mathbf{R})$ is
the number of pairs of atoms separated by $\mathbf{R}$.  In the
derivation above we have utilized the fact that
$\sin{kx_{ab}}=-\sin{kx_{ba}}$ to cancel the sum of
$f(k\mathbf{r}_{ab})\sin{kx_{ab}}$.  In a realistic scenario, one
neither knows $N(\mathbf{R})$ nor $N$ exactly.  In this case, one
must solve instead for the ensemble average $\avg{\delta_p}$ and
the variance $\Delta\delta_p$.  For large $N$, one can safely pull
the factor of $N$ out of the ensemble average:
\be
\avg{\frac{N(\mathbf{R})}{N}}\approx\frac{\avg{N(\mathbf{R})}}{\avg{N}}.
\ee
With this simplification,
\bea \frac{\avg{\delta_p}}{\Gamma} & = &
\frac{1}{2\avg{N}}\cos{2\Omega\tau}\sum^{}_{\mathbf{R}{\neq}0}U(\mathbf{R})\avg{N(\mathbf{R})}
\\ & = &
\frac{1}{2\avg{N}}\cos{2\Omega\tau}\sum^{}_{\mathbf{R}{\neq}0}U(\mathbf{R}){\int}d\mathbf{r}\rho(\mathbf{r})\rho(\mathbf{r}+\mathbf{R})V\label{eq:avgshift},
\\
\frac{(\Delta\delta_p)^2}{\Gamma^2} & = &
\left(\frac{1}{2\avg{N}}\cos{2\Omega\tau}\right)^2\sum^{}_{\mathbf{R},\mathbf{R^\prime}{\neq}0}U(\mathbf{R})U(\mathbf{R}^{\prime})\avg{N(\mathbf{R})N(\mathbf{R}^{\prime})}
\nonumber \\
& &
-\left(\frac{1}{2\avg{N}}\cos{2\Omega\tau}\sum^{}_{\mathbf{R}{\neq}0}U(\mathbf{R})\avg{N(\mathbf{R})}\right)^2
\\ & = &
\left(\frac{1}{2\avg{N}}\cos{2\Omega\tau}\right)^2\left[\sum^{}_{\mathbf{R},\mathbf{R^\prime}{\neq}0}U(\mathbf{R})U(\mathbf{R}^{\prime})
\int{d}\mathbf{r}\rho(\mathbf{r})\rho(\mathbf{r}+\mathbf{R})V^2\Big((1-\rho(\mathbf{r})V)\rho(\mathbf{r}+\mathbf{R^\prime})\right.
\nonumber
\\ & &
+\left.(1-\rho(\mathbf{r}+\mathbf{R})V)\rho(\mathbf{r}+\mathbf{R}+\mathbf{R^\prime})\right.
+\left.(1-\rho(\mathbf{r})V)\rho(\mathbf{r}-\mathbf{R^\prime})\right.
\nonumber
\\ & &
+(1-\rho(\mathbf{r}+\mathbf{R})V)\rho(\mathbf{r}+\mathbf{R}-\mathbf{R^\prime})\Big)
\nonumber
\\ & &
+2\sum^{}_{\mathbf{R}{\neq}0}U(\mathbf{R})^2{\int}d{\mathbf{r}}\rho(\mathbf{r})\rho(\mathbf{r}+\mathbf{R})V\Big((1-\rho(\mathbf{r})\rho(\mathbf{r}+\mathbf{R})V^2)-(1-\rho(\mathbf{r})V)\rho(\mathbf{r}+\mathbf{R})V
\nonumber \\ & &
-\rho(\mathbf{r})(1-\rho(\mathbf{r}+\mathbf{R})V)V\Big)\Bigg]\label{eq:shiftvariance}.
\eea
Each term in Eq.~(\ref{eq:shiftvariance}) has a clear meaning.  To
calculate the average shift in Eq.~(\ref{eq:avgshift}), one must
evaluate $\avg{N(\mathbf{R})}$.
 To do this, one must perform a sum over $\mathbf{r}_i$ of the probability that the sites $\mathbf{r}_i$ and
$\mathbf{r}_i+\mathbf{R}$ are both occupied. To find the variance,
one must calculate quantities like
$\avg{N(\mathbf{R})N(\mathbf{R}^{\prime})}$, and thus the
probability that the sites $\mathbf{r}_i$,
$\mathbf{r}_i+\mathbf{R}$, $\mathbf{r}_j$, and
$\mathbf{r}_j+\mathbf{R}^{\prime}$ are all occupied.  When these
four points are distinct, the probability is simply a product of
the probabilities of each point being occupied. This is untrue
when one or more of the points overlap.  The terms in
Eq.~(\ref{eq:shiftvariance}) represent corrections due to these
overlaps. The product
$\rho(\mathbf{r})\rho(\mathbf{r}+\mathbf{R})V^2((1-\rho(\mathbf{r})V)\rho(\mathbf{r}+\mathbf{R}^{\prime})$,
for example, is due to the overlap of $\mathbf{r}_i$ and
$\mathbf{r}_j$.
\subsection{Effects of multilevel atomic structure}
Our results derived thus far are for the case of two-level atoms.
This is the relevant case of study for the $(J=0)-(J=0)$ forbidden
transition proposed for optical lattice clocks, where the simple
level structure makes it easier to cancel the relative Stark shift
in the clock transition.  Nonetheless, our results can be
generalized to more complicated level structure, such as the case
of an atom with a single ground and multiple excited states. A
simple argument shows that Eq.~(\ref{eq:shift}) remains correct to
the lowest nontrivial order in ${\Gamma}t$.  If multiple excited
states are present in addition to the one that is initially
excited, the equation of motion for
$\avg{\pa{}_{\scriptsize{\textrm{clock}}}}$ in Eq.~(\ref{eq:pa})
will contain additional terms like
$\avg{\za\pb{}_{\scriptsize{\textrm{other}}}}$, where the
subscript ``clock" refers to the clock transition and ``other"
refers to other excited state levels.  Initially,
$\avg{\pb{}_{\scriptsize{\textrm{other}}}}=0$ and thus
\be
\frac{\avg{\za\pb{}_{\scriptsize{\textrm{other}}}}}{\avg{\za\pb{}_{\scriptsize{\textrm{clock}}}}}{\propto}{\Gamma}t.
\ee
Consequently, at short times, evolution of
$\avg{\pa{}_{\scriptsize{\textrm{clock}}}}$ will be dominated by
the clock transition.  Thus, if imperfections in the Ramsey pulse
constitute the major source of shift, the multiple excited states
will contribute an additional source of shift that is first order
in ${\Gamma}t$. If decay of the clock excited state constitutes
the major source, the multiple excited states will contribute a
shift that is of order ${\Gamma}^{2}t^2$.
%
\section{Analysis of results}\label{sec:latticedesign}
Eq.~(\ref{eq:shift}) or Eqs.~(\ref{eq:avgshift})
and~(\ref{eq:shiftvariance}) can be evaluated numerically for a
given lattice configuration and number of atoms. To extract the
key features of the shift, we note that the zeroth-order shift in
$t$ in Eq.~(\ref{eq:shift}) essentially consists of adding
together the classical dipole interaction energies
$\tilde{U}_{ab}{\propto}-\textrm{Re}(\mathbf{d}_{a}{\cdot}\mathbf{E}^{*}_b(\mathbf{r}_a))$.
For a generic configuration of atoms, the amplitudes of the dipole
fields incident on a given dipole $a$ tend to interfere.  For
certain configurations, it is possible that the field amplitudes
will add constructively along some direction of propagation
$\mathbf{k}$. Near these configurations one will expect large
shifts to result. The condition for constructive interference
between radiated dipole fields is similar to that of Bragg
scattering in a crystal, and is readily found to occur when
\be |\mathbf{\tilde{G}}|=k_0, \ee
where $\mathbf{\tilde{G}}=(G_{x}-k_0,G_y,G_z)$ and $\mathbf{G}$ is
a reciprocal lattice vector.  This condition can be rewritten as
\be\label{eq:bigshift} |\mathbf{G}|^2=2k_{0}G_x.\ee
Numerical results indicate that peaks in the line shift do indeed
occur when condition (\ref{eq:bigshift}) is nearly satisfied.

One can easily derive an approximate scaling law for the line
shift in these resonant configurations.  We define a dimensionless
parameter $\beta$ related to the density of atoms by
$n=1/(\beta\lambda)^3$. $\beta$ characterizes the spacing between
neighbors in the lattice.  In a resonant configuration, the
electric fields add constructively, and the total electric field
experienced by an atom is approximately
\be
E\sim{\int}d^3\mathbf{r}\frac{n}{kr}\sim\int_0^{L}dr\frac{nr}{k}\sim\frac{L^2}{k(\beta\lambda)^3},
\ee
where $L$ is the linear size of the system.  For $N$ total atoms,
$L\sim\beta{\lambda}N^{1/3}$.  Then
\be \frac{\delta_p}{\Gamma}\sim\frac{N^{2/3}}{\beta}. \ee

Experimentally, one has freedom to choose the orientations of the
trapping laser beams that form the lattice.  The control
parameters can be parameterized by a set of variables
$\{\alpha\}$, which will also determine the reciprocal lattice
vectors $\mathbf{G}(\{\alpha\})$. One can then find solutions
$\{\alpha_0\}$ of Eq.~(\ref{eq:bigshift}) corresponding to
configurations with large line shifts. In the parameter space
between two sets of solutions $\{\alpha_0\}$, one can numerically
find configurations where the shift is significantly reduced.

In the case of imperfect filling of lattice sites, it will be
important to account for not only the mean shift but the variance
as well. For large numbers of atoms, Eq.~(\ref{eq:shiftvariance})
cannot be evaluated exactly without extensive computational
resources. With a small filling factor $P={\rho}V{\ll}1$, however,
we can estimate that the major contribution to the variance
results from the $\rho^2$ terms, while the $\rho^3$ terms remain
negligible.  In this diffuse limit,
\be\label{eq:diffusevariance}
\frac{(\Delta\delta_p)^2}{\Gamma^2}{\approx}
\left(\frac{1}{2\avg{N}}\cos{2\Omega\tau}\right)^{2}{\times}2\sum^{}_{\mathbf{R}{\neq}0}U(\mathbf{R})^2{\int}d{\mathbf{r}}\rho(\mathbf{r})\rho(\mathbf{r}+\mathbf{R})V.
\ee
In this case, one readily finds that the variance $\Delta\delta_p$
scales like $(P/N)^{1/3}$.  For $P<1/2$, the variance increases
with $P$ due to the increasing uncertainty of whether a pair of
sites will both be occupied, but decreases with $N$ due to the
decreasing fractional uncertainty in the total number of pairs of
atoms separated by a vector $\mathbf{R}$.
%
\section{Numerical example}\label{sec:example}
\begin{figure*}[p]
\begin{center}
\includegraphics[width=9cm]{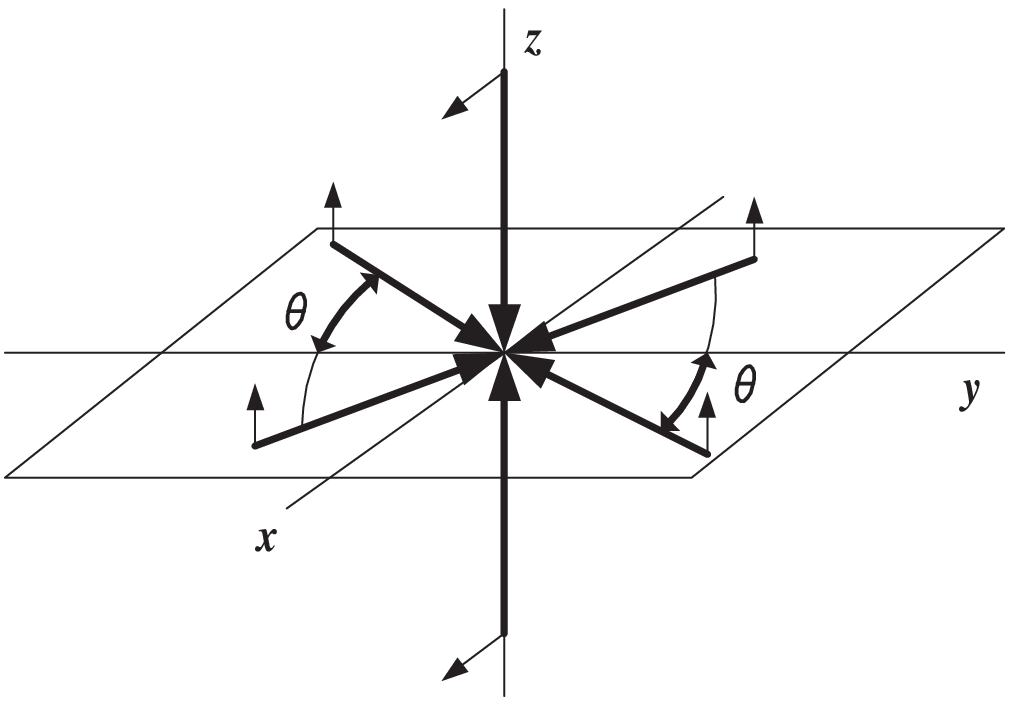}
\end{center}
\caption{The lattice studied in our numerical example is formed by
the interference of six laser beams.  Thick arrows denote the
directions of propagation of the beams, and thin arrows denote the
direction of polarization.  Four beams are oriented along the
$x$-$y$ plane, each making an angle $\pm\theta$ with the $y$-axis
and polarized along $z$. Two additional beams run parallel to $z$
and are polarized along $x$. \label{fig:sixbeam}}
\end{figure*}

\begin{figure*}[p]
\begin{center}
\includegraphics[width=9cm]{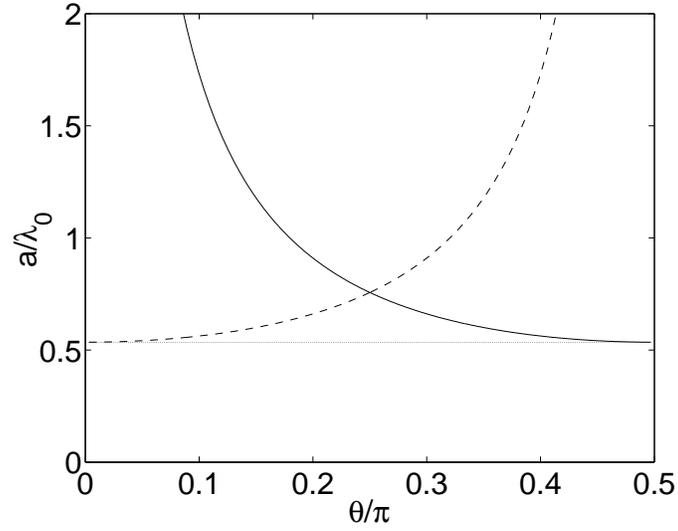}
\end{center}
\caption{The lattice constants for the six-beam lattice are
plotted in units of $\lambda_0$ as functions of $\theta$.  We
assume that the ratio of the resonant wavevector to that of the
trapping lasers is $k_0/k_L=1.07$, consistent with the magic
wavelength of ${}^{87}$Sr. The solid line represents the lattice
constant along $x$, the dashed line along $y$, and the constant
dotted line along $z$.\label{fig:latticeconstants}}
\end{figure*}

As an illustration of our results, we consider ${}^{87}$Sr atoms
trapped in a lattice formed by six interfering beams, as shown in
Fig.~\ref{fig:sixbeam}.  For ${}^{87}$Sr, the ``magic" wavelength
of the trapping lasers is roughly
$\lambda_L=1.07\lambda_0$~\cite{katori02}, and one can vary the
angle $\theta$ between the propagation vectors of the trapping
beams. The resulting lattice is tetragonal, with lattice constants
of $a_x=\pi/k_{L}\sin\theta$, $a_y=\pi/k_{L}\cos\theta$, and
$a_z=\pi/k_{L}$ along $x$, $y$, and $z$, respectively.  The
lattice constants are plotted in Fig.~\ref{fig:latticeconstants}.
The corresponding basis of the reciprocal lattice has lengths
$G_x=2k_{L}\sin\theta$, $G_y=2k_{L}\cos\theta$, and $G_z=2k_{L}$.
We have ignored the effect of atomic
back-action~\cite{deutsch95,weidemuller98} on the trapping fields,
whereby scattering of light by the atoms introduces phases that
might modify the lattice constants.  Such effects are expected to
be stronger in red-detuned lattices, where atoms lie in the
antinodes of the potential, and with increasing atomic density.
Taking into account this back-action does not modify our results,
except that now the lattice constants must be solved
self-consistently~\cite{deutsch95,weidemuller98}.

Using Eq.~(\ref{eq:bigshift}) we can find values of $\theta$ where
constructive interference causes the shifts to be large.  We focus
on two specific solutions, $\theta_0/\pi=0.116$ and
$\theta_0/\pi=0.180$.  For our system we consider $\avg{N}$ atoms
in a spherical distribution with uniform density
$\rho({\mathbf{r}})$ for $r<r_0$, and zero density for $r>r_0$.
The relationship between the density and filling fraction $P$ is
given by $P=\rho({\mathbf{r}})V$, where $V$ is the volume of a
unit cell. The critical value $r_0$ is determined by the equation
\be r_0=\left(\frac{3\avg{N}V}{4{\pi}P}\right)^{1/3}. \ee

\begin{figure*}[p]
\begin{center}
\includegraphics[width=9cm]{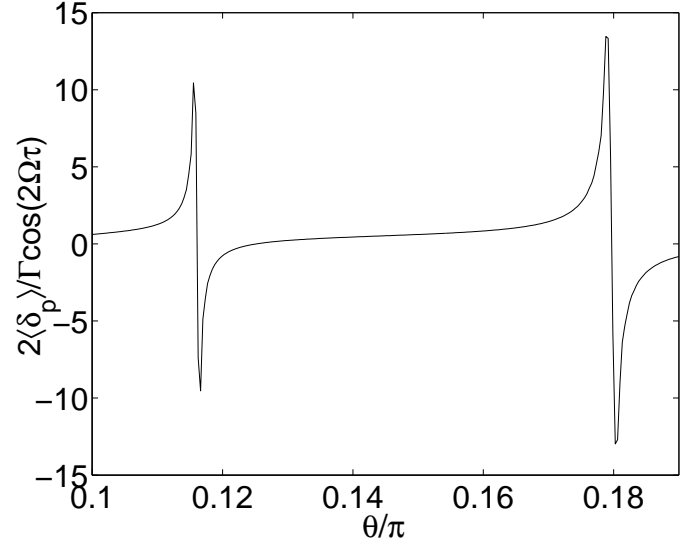}
\end{center}
\caption{The calculated shifts for the six-beam lattice as a
function of $\theta$.  The system consists of $N=10^6$ atoms with
a filling factor of $1$. \label{fig:perfect}}
\end{figure*}

We first consider a perfectly filled lattice consisting of
$N=10^6$ atoms.  For simplicity we calculate the line shift to
zeroth order in the interrogation time $t$.  In
Fig.~(\ref{fig:perfect}), we plot the quantity
$2\delta_p/{\Gamma}\cos(2\Omega\tau)$ as a function of $\theta$.
Peaks in the shift are clearly visible at the points $\theta_0$
that were calculated analytically.  It should be noted that the
line shift can be very large in one of these ``bad"
configurations. Even in the limit of short interrogation times,
one can see that shifts of order $\delta_p{\sim}10\Gamma$ are
possible. This is perhaps a surprising result, and occurs because
the spatial ordering of the atoms allows the interactions to
behave constructively at these points. For longer interrogation
times, one expects this line shift to become even larger, since
the constructive interference in these configurations also leads
to superradiant decay and thus a large contribution to the shift
that is first order in $t$. One also sees that away from these
``bad" points, the shift is strongly suppressed and even becomes
zero for one particular value of $\theta$.

\begin{figure*}[p]
\begin{center}
\includegraphics[width=9cm]{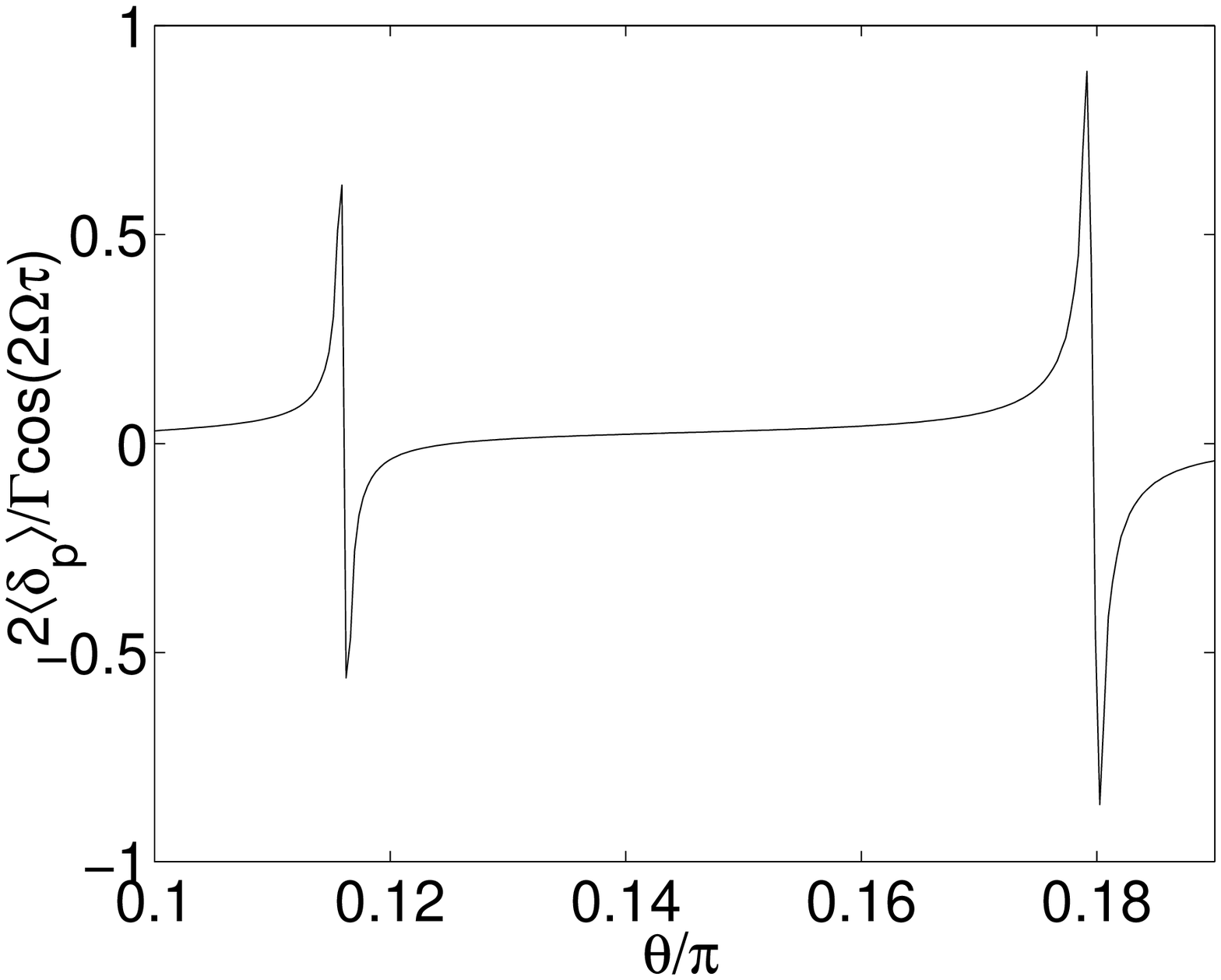}
\end{center}
\caption{The calculated mean shifts for the six-beam lattice as a
function of $\theta$.  The system consists of $\avg{N}=10^5$ atoms
with a filling factor of $0.05$. \label{fig:imperfect}}
\end{figure*}

We next consider a partially filled lattice consisting of
$\avg{N}=10^5$ atoms and a filling factor of $P=0.05$.
Fig.~\ref{fig:imperfect} gives the quantity
$2\avg{\delta_p}/{\Gamma}\cos(2\Omega\tau)$ as a function of
$\theta$. The mean shift $\avg{\delta_p}$ still exhibits peaks at
the points $\theta_0$, and vanishes near $\theta/\pi=0.125$. At
this ``good" point, one can use Eq.~(\ref{eq:diffusevariance}) to
estimate the variance in the expected shift.  Within this diffuse
approximation, we find that the variance
\be
\frac{\Delta\delta_p}{\Gamma}{\approx}\frac{1}{2}\cos(2\Omega\tau){\times}3.1{\times}10^{-3}.
\ee
Experimentally, there will be additional sources of error that
result from not knowing $\rho({\mathbf{r}})$ perfectly, errors in
the configuration of the trapping lasers, and from the effects of
atomic back-action on the lattice constants. Nonetheless, it
appears that the error due to dipole-dipole interactions can be
made quite small by appropriately designing the lattice.

\section{Conclusion}
We have derived an expression for the line shift measured in
Ramsey spectroscopy due to dipole-dipole interactions.  We find
that the lattice geometry strongly affects the magnitude of the
shift, and is peaked in lattice configurations where the
interactions between atoms add constructively.  Because of the
spatial ordering in the lattice, the shift can be quite large in
these resonant configurations. By tuning the lattice between two
of these configurations, one can reduce the dipole-induced line
shift to nearly zero.

While the resonant configurations might be bad for clock
applications, it might be worthwhile to study these configurations
further.  The dipole-dipole couplings in an optical lattice offer
the possibility of strong, constructive interactions that can be
dynamically tuned by changing the lattice geometry. This might be
useful for applications such as quantum information processing and
might have interesting consequences for studying phenomena such as
superradiance and for probing the superfluid-Mott insulator
transition.

\section*{Acknowledgments}
We gratefully acknowledge A. S{\o}rensen, A. Andr{\'{e}}, and R.
Walsworth for many helpful discussions.  This work was supported
by the NSF (CAREER program and Graduate Fellowship), A. Sloan
Foundation, the David and Lucille Packard Foundation, NIST, and
ONR.

\appendix
\section{Derivation of master equation}
In this appendix we derive Eq.~(\ref{eq:dpdt}) starting from the
full atom-field Hamiltonian. For a more detailed derivation and
discussion, one can also see~\cite{agarwal74,gross82}.  The
Hamiltonian for the atom-field system is
\be H=H_0+V, \ee
where
\bea H_0 & = &
H_{\scriptsize{\textrm{internal}}}+H_{\scriptsize{\textrm{field}}}
\\ & = &
\sum^{}_{j}\left(\frac{1}{2}\hbar\omega_0\zj\right)
+\sum^{}_{\mathbf{k},\hat{\epsilon}}\hbar\omega_{\ke}\left(a^{\dagger}_{\ke}a_{\ke}+\frac{1}{2}\right)
\eea
and
\bea V & = & -\sum^{}_{j}\mathbf{d}_{j}\cdot\mathbf{E(\mathbf{r}_j)} \\
& = &
-\sum^{}_{j}\sum^{}_{\ke}d\left(\pj+\mj\right)\left(\hat{\epsilon}_{\scriptsize{\textrm{atom}}}\cdot\hat{\epsilon}\right)
\left(\mathcal{E}_{\ke}a_{\ke}e^{i\mathbf{k}\cdot\mathbf{r}_j}+h.c.\right).
\eea
$\zj$ refers to the internal state of atom $j$, $\omega_0$ is the
atomic transition frequency, and $\omega_{\ke}$ is the frequency
of the mode of the electromagnetic field with wave vector
$\mathbf{k}$ and polarization $\hat\epsilon$. $\mathbf{r}_j$ is
the position of atom $j$, and
$\hat{\epsilon}_{\scriptsize{\textrm{atom}}}$ is the polarization
direction of the dipoles.

We now consider the evolution of the atoms$+$field density matrix
$\rho_{af}$ in the interaction picture.  The equation of motion is
given by
\bea \label{eq:atomfield}\frac{\partial\rho_{af}}{{\partial}t} & =
&
\frac{1}{i\hbar}\left[\tilde{V}(t),\rho_{af}\right], \\
\tilde{V}(t) & = & e^{iH_{0}t/\hbar}Ve^{-iH_{0}t/\hbar}. \eea
We can integrate Eq.~(\ref{eq:atomfield}) once and substitute the
result back into itself.  We then trace out the field degrees of
freedom to obtain an equation of motion for the atomic density
matrix $\rho$ alone:
\be \frac{\partial\rho}{{\partial}t} =
-\frac{1}{\hbar^2}\textrm{Tr}_{f}\int^{t}_{0}d{\tau}\left[\tilde{V}(t),
\left[\tilde{V}(t-\tau),\rho_{af}(t-\tau)\right]\right]. \ee
To make the above equation more useful, we employ the Born-Markov
approximation, replacing $\rho_{af}(t-\tau)$ with
$\rho(t)\otimes\ket{0}\bra{0}$.  Physically, this amounts to
assuming that correlations in the field are negligible, that the
field can always be approximated by a vacuum state, and that the
correlation time of the atom-field system is much shorter than any
atomic timescales.  These assumptions safely allow us to extend
the time integral to infinity, so that
\be\label{eq:bornmarkov} \frac{\partial\rho}{{\partial}t} =
-\frac{1}{\hbar^2}\textrm{Tr}_{f}\int^{\infty}_{0}d{\tau}\left[\tilde{V}(t),
\left[\tilde{V}(t-\tau),\rho(t)\otimes\ket{0}\bra{0}\right]\right].
\ee

Writing out $\tilde{V}(t)$ and $\tilde{V}(t^{\prime})$, where
$t^{\prime}=t-\tau$, give
\bea \tilde{V}(t) & = & \sum^{}_{a,\mathbf{k},\hat{\epsilon}}
-d\mathcal{E}_{\ke}\left(\hat{\epsilon}_{\scriptsize{\textrm{atom}}}\cdot\hat{\epsilon}\right)
\left({\pa}e^{i\omega_{0}t}+{\ma}e^{-i\omega_{0}t}\right)
\left(a_{\ke}e^{i(\mathbf{k}{\cdot}\mathbf{r}_a-{\omega}_{\ke}t)}
+
a^{\dagger}_{\ke}e^{-i(\mathbf{k}{\cdot}\mathbf{r}_a-{\omega}_{\ke}t)}\right),
\\
\tilde{V}(t^{\prime}) & = &
\sum^{}_{b,\mathbf{k^{\prime}},\hat{\epsilon}^{\prime}}
-d\mathcal{E}_{\kep}\left(\hat{\epsilon}_{\scriptsize{\textrm{atom}}}\cdot\hat{\epsilon}^{\prime}\right)
\left({\pb}e^{i\omega_{0}t^{\prime}}+{\mb}e^{-i\omega_{0}t^{\prime}}\right){\times}
\nonumber \\
& &
\left(a_{\kep}e^{i(\mathbf{k}^{\prime}{\cdot}\mathbf{r}_b-{\omega}_{\kep}t^{\prime})}
+
a^{\dagger}_{\kep}e^{-i(\mathbf{k}^{\prime}{\cdot}\mathbf{r}_b-{\omega}_{\kep}t^{\prime})}\right).
\eea
When we substitute the expansions above into
Eq.~(\ref{eq:bornmarkov}) and perform the trace, the only nonzero
terms will be those associated with
$a_{\ke}a^{\dagger}_{\kep}\ket{0}\bra{0}$,
$a^{\dagger}_{\ke}\ket{0}\bra{0}a_{\kep}$,
$a^{\dagger}_{\kep}\ket{0}\bra{0}a_{\ke}$, and
$\ket{0}\bra{0}a_{\kep}a^{\dagger}_{\ke}$, where
$\mathbf{k}=\mathbf{k}^{\prime}$ and
$\hat{\epsilon}=\hat{\epsilon}^{\prime}$.  Making these
simplifications and replacing the sum on $\mathbf{k}$ by an
integral gives
\bea \frac{\partial\rho}{{\partial}t} & = &
-\frac{1}{\hbar^2}\int^{\infty}_{0}d{\tau}\int^{\infty}_{0}\frac{1}{(2\pi)^3}k^{2}\,dk{\int}d\Omega\sum^{}_{a,b,\hat{\epsilon}}d^{2}
\mathcal{E}_{\ke}^{2}\left(\hat{\epsilon}_{\scriptsize{\textrm{atom}}}\cdot\hat{\epsilon}\right)^{2}
\left[e^{-i(\omega_{\ke}-\omega_0)\tau+i\mathbf{k}\cdot\mathbf{r}_{ab}}\pa\mb\rho\right.
\nonumber \\ & &
\left.+e^{-i(\omega_{\ke}+\omega_0)\tau+i\mathbf{k}\cdot\mathbf{r}_{ab}}\ma\pb\rho
-e^{i(\omega_{\ke}+\omega_0)\tau-i\mathbf{k}\cdot\mathbf{r}_{ab}}\pa\rho\mb\right.
\nonumber
\\ & & \left.-e^{i(\omega_{\ke}-\omega_0)\tau-i\mathbf{k}\cdot\mathbf{r}_{ab}}\mb\rho\pa
+h.c.\right]. \eea
The angular integral is tedious but straightforward and gives
\be
{\int}d\Omega\sum^{}_{\hat{\epsilon}}e^{i\mathbf{k}\cdot\mathbf{r}_{ab}}\left(\hat{\epsilon}_{\scriptsize{\textrm{atom}}}\cdot\hat{\epsilon}\right)^{2}=
4\pi\left(\sin^{2}\theta\frac{\sin{kr_{ab}}}{kr_{ab}}+\left(1-3\cos^{2}\theta\right)\left(\frac{\cos{kr_{ab}}}{(kr_{ab})^2}-\frac{\sin{kr_{ab}}}{(kr_{ab})^3}\right)\right).
\ee
The time integral can be evaluated using the formula
\be
\int^{\infty}_{0}c\,d{\tau}e^{-ic(k{\mp}k_0)\tau}={\pi}\delta(k_0{\mp}k){\pm}i\mathcal{P}\frac{1}{k_{0}{\mp}k},
\ee
where $\mathcal{P}$ denotes the principal value.  The delta
function above eventually yields the non-Hermitian component of
the evolution, while the principal value yields the coherent
atom-atom interactions.  Performing all integrals, making the
replacement $\Gamma=k_{0}^{3}d^2/3\pi\epsilon_{0}{\hbar}c^3$, and
changing back to the Schrodinger picture yields
Eq.~(\ref{eq:dpdt}).
\section{Second quantization of atoms in lattice}
The results above were derived treating the atoms in the lattice
in first quantization.  Starting instead from second quantization,
one can see that the results derived in first quantization are
appropriate in the limit of tight confinement of the atoms, when
the overlap between atoms at different sites can be ignored.

In second quantization, the atomic wave function
$\hat{\psi}(\mathbf{r})$ can be expanded in terms of Wannier
functions:
\be
\hat{\psi}(\mathbf{r})=\sum^{}_{{\nu}ni}b_{{\nu}ni}\phi_{{\nu}n}(\mathbf{r}-\mathbf{r}_i).
\ee
Here $\nu=e,g$ denotes the internal state of the atom, $n$ is the
band index, and $i$ labels the lattice sites $\mathbf{r}_i$.  To
lowest order, we assume that only the harmonic oscillator
ground-state wave function is relevant.  The atomic Hamiltonian is
then given in second quantization by
\be
\hat{H}_a=\sum^{}_{{\nu}ij}b_{i\nu}^{\dagger}b_{j\nu}\int{d\mathbf{r}}\phi_{\nu}(\mathbf{r}-\mathbf{r}_i)
(\frac{\hat{P}^2}{2M}+U(\mathbf{r})+E_{\nu})\phi_{\nu}(\mathbf{r}-\mathbf{r}_j).
\ee
In the limit of tight confinement, the overlap integrals for
$i{\neq}j$ can be ignored, leading back to the atomic Hamiltonian
used in first quantization.

In second quantization, the electric dipole Hamiltonian is given
by
\be
H_e=-\sum^{}_{ij}\left(b_{ie}^{\dagger}b_{jg}\int{d\mathbf{r}}\phi_{e}(\mathbf{r}-\mathbf{r}_i)(\mathbf{d}\cdot\mathbf{E}(\mathbf{r}))\phi_{g}(\mathbf{r}-\mathbf{r}_j)
+b_{ig}^{\dagger}b_{je}\int{d\mathbf{r}}\phi_{g}(\mathbf{r}-\mathbf{r}_i)(\mathbf{d}\cdot\mathbf{E}(\mathbf{r}))\phi_{e}(\mathbf{r}-\mathbf{r}_j)\right),
\ee
where
\be
\mathbf{E}(\mathbf{r})=\sum^{}_{\ke}\mathcal{E}_{\ke}a_{\ke}e^{i\mathbf{k}\cdot\mathbf{r}}\hat{\epsilon}+h.c.
\ee

Again, in the limit of tight confinement, the overlap integrals
for $i{\neq}j$ can be ignored, and furthermore the term
$\mathbf{E}(\mathbf{r})$ can be replaced with
$\mathbf{E}(\mathbf{r}_i)$.  In this limit this Hamiltonian is
equivalent to the electric dipole Hamiltonian used in first
quantization.

%

\end{document}